\documentclass[12pt,aps,floats,prc]{revtex4}
\usepackage{graphicx}
\usepackage[dvips]{color}
\usepackage{colordvi}  
\def\Journal#1#2#3#4{{\em #1} {\bf #2}, #3 (#4) }
\def\NPA{{ Nucl. Phys.} A}

\def\PRC{{Phys. Rev.} C}

\def\EJA {Eur. Phys. J. A}

\begin{document}
\title{Spectral function at high missing
energies and momenta}
\author{T. Frick$^1$, Kh.S.A. Hassaneen$^1$, D. Rohe$^2$,  and H. M\"uther$^1$}
\affiliation{$^1$ Institut f\"ur
Theoretische Physik, \\ Universit\"at T\"ubingen, D-72076 T\"ubingen, Germany}
\affiliation{$^2$ Department of Physics and Astronomy, \\ University of Basel,
CH-4056 Basel, Switzerland}
\begin{abstract}
The nuclear spectral function at high missing energies and momenta has been
determined from a self-consistent calculation of the Green's function in nuclear
matter using realistic nucleon-nucleon interactions. The results are compared 
with recent experimental data derived from ($e,e'p$) reactions on $^{12}C$. A
rather good agreement is obtained if the Green's functions are calculated in a
non-perturbative way.
\end{abstract}
\maketitle

\section{Introduction\label{Introduction}}
The nuclear many-body problem is one of the challenging problems for quantum
many-body theory. The strong components of a realistic nucleon-nucleon (NN)
interaction, in particular the short-range and tensor components, induce strong
correlations in the nuclear wave function. In fact, a calculation of the binding
energy of nuclei, which ignores the effects of correlations beyond the
Hartree-Fock or mean-field approximation, yields unbound nuclei, 
if realistic NN interactions are considered\cite{her1}. 

The investigation of such correlations should provide information about the
interactions of two nucleons in the nuclear medium.  Therefore considerable
experimental as well as theoretical effort has been made to explore these
correlations and determine their features from observables which are sensitive
to correlation effects. Since one wants to investigate correlations beyond the
mean field approximation, it is plausible that one tries to determine the
probability for nucleons occupying states which are unoccupied in the
Independent Particle Shell Model (IPSM). This implies that one tries to detect
the probability of nucleons with momenta considerably above the Fermi momentum.

Attempts have been made to investigate such high-momentum components in
exclusive $(e,e'p)$ experiments, for which the energy transfer, determined 
from the 
 energy of the electron and the energy of the knocked-out proton,
are constrained to guarantee that the residual nucleus is in its groundstate.
Using the Plane-Wave Impuls Approximation (PWIA) the momentum of the nucleon
before absorbing the virtual photon can be derived from the momentum transfer
from the electrons and the momentum of the outgoing proton. This means that one
tries to measure the probability to remove a nucleon carrying large momentum,
depositing only little energy. This probability is given by the spectral
function at high momenta and small missing energies.

The experimental data on the spectral function at high momenta and small missing
energies can very well be described in terms of momentum distributions derived
from IPSM calculations if the theoretical strengths are quenched by
a global spectroscopic factor $Z$\cite{blom,bobel,exp1}. A more detailed analysis of
the spectral function showed that correlations should show up in an enhancement
of the spectral function, as compared to IPSM, at high momenta {\em and} large missing
energies\cite{herwim1,her2}. 

Due to the small cross section  in this region as well as the distortion from
other reaction mechanisms this  requires more experimental
effort\cite{rohe,rohe1}. 
A calculation of the contribution from  rescattering \cite{Barb03b} of the
knocked--out proton in the nucleus confirms that parallel  kinematics has to be
favoured compared to the perpendicular one. The correction on the  spectral
function measured for $^{12}C$ at high missing momentum due to rescattering is
negligible.

In this manuscript we would like to study results for the spectral function at
high missing momenta and energies, which are deduced from those recent
($e,e'p$) experiments\cite{rohe,rohe1}, with calculations within the framework
of the self-consistent Green's function
approach\cite{her1,bal1,dick2,wimn,boz2}. Special attention is paid to the
questions of the necessary ingredients for a calculation to provide
reliable predictions for the spectral function at the energies and momenta of
interest.

After this introduction we present a short review on the extraction of the
spectral function  in section 2. A comparison of results deduced from various
approximations schemes and the experimental data is made in section 3, while
the last section contains a summary and the conclusions.

\section{Spectral function }      

\begin{figure}
\begin{center}
\includegraphics[scale=0.8,angle=0]{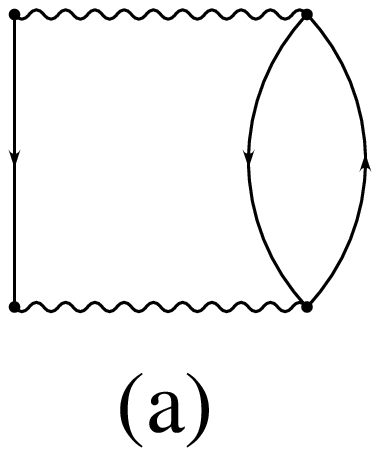}
\includegraphics[scale=0.8,angle=0]{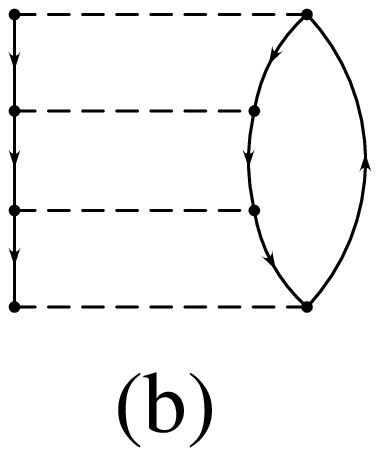}
\includegraphics[scale=0.8,angle=0]{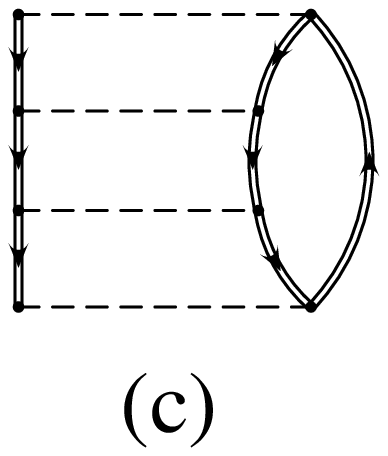}
\begin{center}
\caption{\label{fig_diags} Time ordered diagrams included in the
  different approximations for the self energy. In the second-order
  perturbative 
  approximation (a), the bare potential (dashed line) is replaced 
  by a Brueckner $G$~matrix (wiggly line). The quasiparticle approximation
  includes hole-hole scattering of quasiparticles to all orders (b), and the
  self-consistent approximation includes hole-hole scattering
  also off-shell (c).}
\end{center}
\end{center}
\end{figure}

Within the framework of the self-consistent evaluation of Green's functions
the spectral function in nuclear matter is calculated from the
retarded nuclear self energy, 
\begin{equation}
\label{spec_function} 
A(k,\omega)=\frac{-2\,{\mathrm{Im}}\,\Sigma(k,\omega+{\mathrm{i}}\eta)}
{[\omega-\frac{k^2}{2m}-{\mathrm{Re}}\,\Sigma(k,\omega)]^2+
[{\mathrm{Im}}\,\Sigma(k,\omega+{\mathrm{i}}\eta)]^2}.
\end{equation}
Different approximation schemes can be chosen for the self energy.  A first
method to approximate the self energy in the energy domain that can be explored
in knock-out experiments is to add a  perturbative  two-hole-one-particle
contribution to the standard Brueckner self energy, 
\begin{equation}
\label{eq_2h1p}
\Delta \Sigma^{2h1p}(k,\omega+{\mathrm{i}}\eta)=
\int_{k_F}^\infty\frac{{\mathrm{d}}^3k^{\prime}}{(2\pi)^3}
\int_0^{k_F}\frac{{\mathrm{d}}^3h}{(2\pi)^3}
\int_0^{k_F}\frac{{\mathrm{d}}^3h^{\prime}}{(2\pi)^3} \frac{
\left<{\mathbf{k}}{\mathbf{h}}\right|G
\left|{\mathbf{k}}^{\prime}{\mathbf{h}}^{\prime}\right>_A^2
}{\omega+\epsilon_{k^{\prime}}-\epsilon_h-\epsilon_{h^{\prime}}+{\mathrm{i}}\eta}
\end{equation} 
This expression is given by the  second order diagram shown in
Fig.~\ref{fig_diags} (a). Instead of the bare potential $V$, 
expression~(\ref{eq_2h1p}) contains an effective interaction, the Brueckner
$G$~matrix.  The single-particle energies $\epsilon_{k^{\prime}}$ are taken
from a Brueckner-Hartree-Fock spectrum.  

A more sophisticated ansatz for the self energy is the ladder approximation.
However, it turned out that this ladder approximation is plagued by the
so-called pairing instability\cite{ram1}. To
circumvent the pairing instability that makes calculations impossible in the
low-density system,  one can use the theory of finite-temperature
Green's functions. Above the transition temperature to the superfluid state,
the pairing instability disappears. For low temperatures, up to
$T=5\,\mbox{MeV}$, the modifications that are induced by the thermal motion
will be restricted to a small energy interval around the chemical potential
$\mu$.\\

In the framework of self-consistent Green's functions (SCGF), the
imaginary part of the ladder self energy $\Sigma^L$ 
is computed from an effective two-nucleon interaction, the nuclear 
$T$~matrix,
\begin{eqnarray}
\label{eq_sigma_aftermatsubara}
{\mathrm{Im}}\Sigma^L(k,\omega+{\mathrm{i}}\eta)
&=&
\int\frac{{\mathrm{d}}^3k^{\prime}}{(2\pi)^3}
\int_{-\infty}^{+\infty} \frac{{\mathrm{d}}\omega^{\prime}}{2\pi}
\left<{\mathbf{kk}}^{\prime}|
{\mathrm{Im}}T(\omega+\omega^{\prime}+{\mathrm{i}}\eta)|
{\mathbf{kk}}^{\prime}\right>_A\nonumber\\
&&\times
A(k^{\prime},\omega^{\prime})
[f(\omega^{\prime})+b(\omega+\omega^{\prime})].
\end{eqnarray}
Here, $b(\Omega)=[e^{\beta(\Omega-2\mu)}-1]^{-1}$ is the Bose function and
$\beta=1/T$ the inverse temperature. 
The real part of the self energy can be obtained as the sum of a
Hartree-Fock contribution and an energy dependent term that is
computed from the imaginary
part by means of a dispersion integral~\cite{fri03}. 
The effective interaction is determined by the following integral equation,
\begin{eqnarray}
\label{eq_twaves}
\left<q|T^{JST}_{LL^{\prime}}(P,\Omega+{\mathrm{i}}\eta)|q^{\prime}\right>
&=&
\left<q|V^{JST}_{LL^{\prime}}|q^{\prime}\right>
+\sum_{L^{\prime \prime}} \int_0^{\infty} 
\frac{
{\mathrm{d}} k^{\prime}\,k^{\prime 2}
}
{(2\pi)^3}
\left<q|V^{JST}_{LL^{\prime \prime}}|k^{\prime}\right>\nonumber\\
&&
\times
{\bar{g}}_{\mathrm{II}}(P,\Omega+{\mathrm{i}}\eta,k^{\prime})
  \left<k^{\prime}|T^{JST}_{L^{\prime \prime}L^{\prime}}(P,\Omega+{\mathrm{i}}
\eta)|q^{\prime}\right>.
\end{eqnarray}
By means of the non-interacting two-particle propagator $g_{\mathrm{II}}$, 
the $T$~matrix takes into account multiple off-shell scattering of particle
pairs and hole pairs to all orders, 
\begin{equation}
\label{eq_pp_and_hh}
g_{\mathrm{II}}(k_1,k_2,\Omega+{\mathrm{i}}\eta)=
\int_{-\infty}^{+\infty}\frac{{\mathrm{d}}\omega}{2\pi}
\int_{-\infty}^{+\infty}\frac{{\mathrm{d}}\omega^{\prime}}{2\pi}
A(k_1,\omega)A(k_2,\omega^{\prime})
\frac{1-f(\omega)-f(\omega^{\prime})}
{\Omega-\omega-\omega^{\prime}+{\mathrm{i}}\eta}.
\end{equation}
To be used in the integral equation~(\ref{eq_twaves}),
$g_{\mathrm{II}}$ must be 
averaged over the angle between the center of mass
momentum ${\mathbf{P}}=\frac{1}{2}({\mathbf{k}}_1+{\mathbf{k}}_2)$ and
the relative momentum
${\mathbf{q}}=\frac{1}{2}({\mathbf{k}}_1-{\mathbf{k}}_2)$.\\
The particles can be forced to propagate on the energy shell by
introducing a quasiparticle spectrum
\begin{equation}
\label{qp_energy}
\epsilon(k)=\frac{k^2}{2m}+{\mathrm{Re}}\Sigma(k,\epsilon(k)),
\end{equation} 
which yields a Dirac type spectral function,
\begin{equation}
\label{eq_sf}
A^{QP}(k,\omega)=2\pi\,\delta(\omega-\epsilon(k)),
\end{equation}   
and a simple representation for $g_{\mathrm{II}}$ in terms of
the spectrum,
\begin{equation}
\label{eq_qp_gII}
g^{QP}_{\mathrm{II}}(k_1,k_2,\Omega+{\mathrm{i}}\eta)=
\frac{
1-f(\epsilon(k_1))-f(\epsilon(k_2))
}
{\Omega-
\epsilon(k_1)+\epsilon(k_2)
+{\mathrm{i}}\eta}.
\end{equation}
Solving for the ladder self energy $\Sigma^{QP}$ and the $T$~matrix 
in the simplified scheme by applying
Eqs.~(\ref{eq_sf}) and~(\ref{eq_qp_gII}), still, all ladders are taken
into account, however, the self-consistency constraint is only
fulfilled on-shell.\\

Since we want to discuss the influence of correlation on proton knock-out
experiments, we focus our attention to the spectral function $A(k,\omega )$ and
the self-energy $\Sigma (k,\omega )$ at energies $\omega$ below the Fermi energy
or chemical potential $\mu$. The time ordered diagrams presented in
Fig.~\ref{fig_diags} are typical contributions to the self-energy in this 
energy region, representing the various approximations, which we are going to
discuss. As already mentioned before, the diagram of Fig.~\ref{fig_diags} (a)
represents the contribution to the self-energy of second order in the Brueckner
$G$-matrix. The propagation of the intermediate two-hole-one-particle state is
described in terms of the quasiparticle approximation to the Green's function.
It represents the contribution of Eq.~(\ref{eq_2h1p}) and we will call the
approximation, which accounts for this term only, the perturbative approach.

The quasiparticle approximation goes beyond this perturbative approach in the
sense that it accounts for all particle-particle hole-hole ladder diagrams, like
the one displayed in Fig.~\ref{fig_diags} (b). The name quasiparticle 
approximation refers to the fact, that also in this approximation, the
single-particle Green's functions, which enter the definition of the
self-energy, are approximated by the quasiparticle approach (\ref{eq_sf}).

Finally, the SCGF approximation refers to calculations, in which the
self-energy (\ref{eq_sigma_aftermatsubara} ) and the corresponding integral 
equation for $T$ (\ref{eq_twaves} ) are
solved considering single-particle propagators, which account for the complete
spectral function. This is represented by the double lines in the diagram of 
Fig.~\ref{fig_diags} (c). One may consider these double lines to represent a
coupling of the propagation of particles (holes) to two-particle-one-hole
(two-hole-one-particle) configurations. Therefore the diagram displayed in 
Fig.~\ref{fig_diags} (c) implies the admixture of ($n+1$)-hole-$n$-particle
configurations in the definition of the self-energy.

\section{Results and discussion}
The approximations (perturbative -, quasiparticle - and
SCGF approximation) just described have been employed in studies of nuclear
matter at a density $\rho$ which is one half of the empirical saturation density
of nuclear matter. We have chosen this density as we want to compare the results
for spectral functions with those obtained in ($e,e'p$) experiments on $^{12}$C
and this density corresponds to the mean value of the particle density of
this nucleus. The assumption that the spectral function of a finite nucleus can
be compared to one for infinite nuclear matter is certainly not justified when
considering the spectral function at low momenta and energies. In this regime the
spectral function is dominated by the single-particle wave functions, which are
of course quite different for infinite matter and finite nuclei. 
Our comparison shall focus on the spectral function at high momenta and high
missing energies. The spectral function in this region is dominated by the
short-range correlations, which should not be so sensitive to the global
properties of the system.

Self-consistent calculations have been performed using the three approximation
schemes in the sense, that for each approach we calculated the quasiparticle
spectrum or the spectral functions, which enter the evaluation of the 
self-energy, from the corresponding self-energy. The CD-Bonn
interaction\cite{cdbonn} has been used for the NN interaction.

\begin{figure}
\begin{center}
\center\includegraphics[scale=0.45]{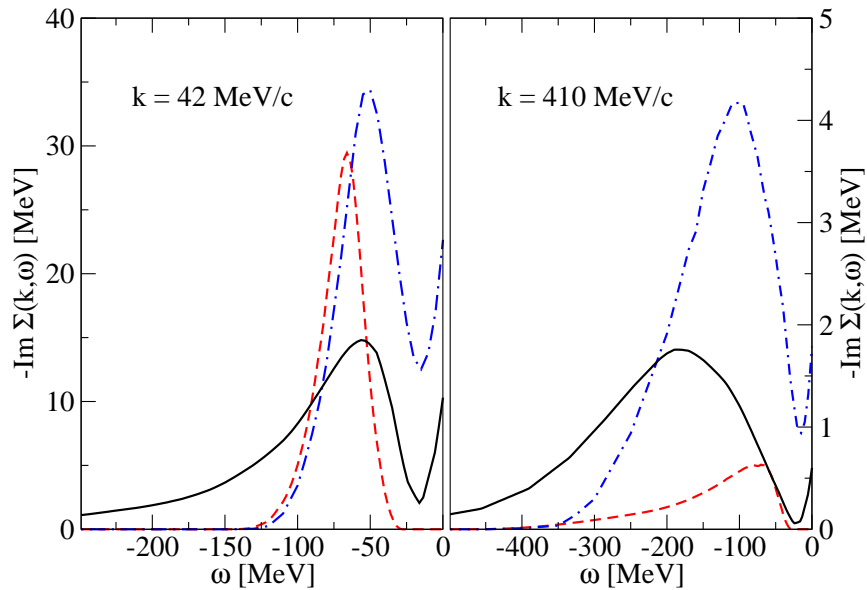}
\end{center}
\caption{\label{fig_se}(Color online) Imaginary part of the nuclear self energy
in nuclear matter at $0.5\rho_0$. The perturbative 2h1p result (dashed line) is
compared to the quasiparticle (QPGF) approximation (dash-dotted line) and the
full SCGF result (solid line). The momentum is $0.2k_F$ in the left panel and
about $2k_F$ in the right panel.}
\end{figure}

As a first step it 
is instructive to compare the three different approximation schemes
on the level of the self energy. In Fig.~\ref{fig_se}, the imaginary
part of $\Sigma$ is displayed for a small momentum that corresponds to
$k_F/5$ and a large momentum above $2k_F$.  As explained above, the
calculations are performed at different temperatures, $T=0$ for the
perturbative approach, $T=2\,\mbox{MeV}$ for the SCGF solution and
$T=5\,\mbox{MeV}$ for the quasiparticle approximation, the latter
displaying a higher transition temperature to the superfluid state.
Due to the finite temperature, the self energy does not go to zero at
the chemical potential, which is between $-20$ and 
$-30\,\mbox{MeV}$ in the quasiparticle and the SCGF
calculations. 

At low momenta (left panel Fig.~\ref{fig_se}) the self-energy of the
perturbative approach,  ${\mathrm{Im}}\Sigma^{2h1p}$, and the quasiparticle
approximation, ${\mathrm{Im}}\Sigma^{QP}$, exhibit a rather similar energy
dependence in the tail towards negative energies, i.e.~high missing
energies. These two approximations also yield well pronounced maxima at 
 around $30\,\mbox{MeV}$. The result for the SCGF approach is quite
different. The maximum is reduced and the tail extends to larger missing
energies. This can be understood from the fact that the SCGF approach accounts
for the complete spectral function (with tails to large energies) already in 
calculating the self-energy, while these contributions are missing in the
perturbative and the quasiparticle approximation. In other words: the SCGF
approach accounts for the admixture of 3-hole-2-particle, 4-hole-3-particle 
etc. configurations, which are missing in the other approximations, and
therefore yields an imaginary part for the self-energy and spectral strength 
also at higher missing energies.

This feature can also be observed in the right panel of Fig.~\ref{fig_se},
displaying the spectral function at larger momenta. In this case one furthermore
observes a significant difference  between the perturbative and the
quasiparticle approximation: the imaginary part resulting from the perturbative
calculation is much weaker than the one obtained in the calculations accounting
for the particle-particle hole-hole ladders to all orders. This is a clear
indication that the nonperturbative  determination of the correlations is
required to obtain a reliable information about the imaginary part of the
self-energy or the spectral function at large momenta. Note that the scale in
the right panel of Fig.~\ref{fig_se} is reduced as compared to the left one.
   
\begin{figure}
\begin{center}
\center\includegraphics[scale=0.45]{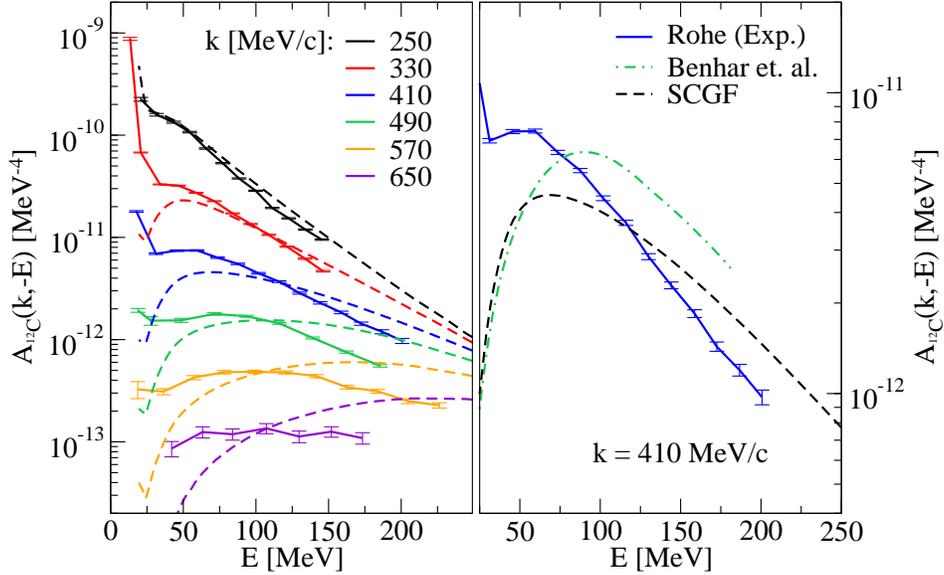}
\end{center}
\caption{\label{fig_rohe}(Color online) Spectral function  
in the $^{12}{\mathrm{C}}$ nucleus. 
Left panel: Experimental result for several momenta above the Fermi
momentum (solid lines with error bars). 
The data are similar to those presented in Ref.~\protect{\cite{rohe}}, but
the choice $cc$ was used for $\sigma_{ep}$ instead of $cc1$.
The dashed lines represent the SCGF nuclear matter spectral 
function at a density of $\rho=0.08\,\mbox{fm}^{-3}$. 
To compare the nuclear matter spectral function with the
experimental result, it must be multiplied by a normalization factor
of~$4Z(2\pi)^{-4}\rho^{-1}$. 
Right panel: A comparison between the experimental result at 
$k=410\,\mbox{MeV}$ (solid lines with error bars), the
theoretical spectral functions for a finite system  
by Benhar {\it et.~al.}~\protect{\cite{ben94}} (dashed line) 
and the SCGF result (solid line).}
\end{figure}

The measured spectral function in the left panel of
Fig.~\ref{fig_rohe} provide direct 
evidence for the partial occupation of single
particle states above the Fermi level by off-shell particles. 
The nuclear matter spectral function 
derived in the theory of SCGF is
compared to the experimental spectral functions 
for different momenta above $k_F=209\,\mbox{MeV/c}$. The prediction derived from
the calculation in nuclear matter should not be reliable at low missing
energies. At these energies the spectral function should be influenced by
long-range correlations, which are different in infinite matter as compared to
finite nuclei.
The agreement is rather good up to $2k_F$. For even higher momenta, 
the energy
dependence is not well reproduced and the theoretical result tends to
overshoot at high missing energies whereas for smaller energies strength is
missing.

The right panel of Fig.~\ref{fig_rohe} shows different theoretical
calculations for a specific nucleon momentum of $k=410\,\mbox{MeV/c}$. 
Compared to the data the spectral function from the CBF theory has the same 
tendency as the SCGF approach described above. However, the SCGF approach 
seems to be closer to the experimental results at high missing energies.

Further, one can compare the correlated strength from the SCGF approach to the 
experimental result as it was done for the CBF theory \cite{roh04}. For this
purpose we integrate the  spectral function over momentum and energy in the
correlated  region above the Fermi level covered by the experiment. The lower
limit of the energy  in the integral has been taken as 40~MeV to avoid the
contribution of the  single--particle levels. The correlated strength from the
SCGF theory in the  integration region specified in \cite{roh04} is 0.61. The
experimental value is  0.61 $\pm$ 0.06, which contains a contribution from
rescattering of 4~\%. Good  agreement is achieved.   

\section{Conclusions}
The study of the spectral function for nucleon knock-out experiments at high
momenta and large missing energies seems to be an appropriate tool to explore
the effect of correlations on the nuclear many-body wave function. Our study
demonstrate that non-perturbative calculations are required to predict the
spectral strength at high momenta. Furthermore one has to account for the
admixture of configurations beyond two-hole-one-particle to obtain a reliable
prediction for the energy tail of the spectral functions towards large missing
energies. These ingredients are taken into account in the SCGF (self-consistent 
Green's functions) approach. The comparison of the spectral function derived
from experimental data with the results obtained from nuclear matter calculation
using SCGF indicates that the effects of short-range correlations are
insensitive to the bulk structure of the nuclear system. The study of the
spectral function at low missing energies will require a more detailed
investigation of long-range correlation and should be performed for the finite
system. 

The authors would like to thank Ingo Sick for many discussions. Also 
we would like to acknowledge financial support from the {\it Europ\"aische
Graduiertenkolleg T\"ubingen - Basel}, which is supported by the {\it Deutsche
Forschungsgemeinschaft} (DFG) and the {\it Schweizerische Nationalfonds} (SNF).

\end{document}